\documentclass[12pt,prd]{revtex4}
\usepackage[cp1251]{inputenc}
\usepackage[english,russian]{babel}
\usepackage{epsf}
\usepackage[dvips]{graphicx}
\begin{document}
\baselineskip =0.8\baselineskip
\vskip 1cm


%
%

\title
{
Self-similarity of negative particle production \\
from the Beam Energy Scan Program at STAR
}

\author{Mikhail Tokarev \\{(on behalf of STAR Collaboration)}}

\address{Joint Institute for Nuclear Research, \\
Joliot Curie 6, Dubna, Moscow region, 141980,
Russia
\\
tokarev@jinr.ru}



\begin{center}
{
\large \bfseries
 Self-similarity of negative particle production \\
from the Beam Energy Scan Program at STAR
}
\vskip 5mm
 M.V.~Tokarev \\ 
{(on behalf of STAR Collaboration)}\\
{{\it Joint Institute for
Nuclear Research, Dubna, Russia}} \\[5mm]

\vskip 5mm

{\it E-mail: tokarev@jinr.ru}
\end{center}

\vskip 5mm
\centerline{\bf Abstract}
We present the spectra of negative charged particle production in Au+Au collisions
from STAR for the first phase of the RHIC Beam Energy Scan Program 
measured  over a wide range 
of collision energy $\sqrt {s_{NN}}$=7.7-200~GeV,
 and transverse momentum of produced particle
in different centralities at $|\eta|<0.5$.
The spectra demonstrate strong dependence on 
collision energy which enhances with $p_T$. 
An indication of self-similarity of negative charged particle 
production in Au+Au collisions is found.  
The constituent energy  loss as a function of energy and centrality 
of collisions and transverse momentum  of inclusive particle 
was  estimated  in the $z$-scaling approach.
The energy dependence of the model parameters - the fractal 
and fragmentation dimensions and "specific heat",  was studied.
\\[10mm]


\vskip 3cm

\begin{center}
Presented at the International Conference "HADRON STRUCTURE'15" \ , \\ 
29 June — 3 July, 2015, Horn\'{y} Smokovec, Slovak Republic. 
\end{center}

\newpage

\section{Introduction}

A major goal of high-energy heavy-ion collision experiments at RHIC 
is to determine the phase diagram for nuclear matter and describe 
its features in framework  of the Quantum Chromodynamics (QCD). 
The most experimentally accessible way to characterize 
the QCD phase diagram is in the plane of temperature $(T)$ and the baryon 
chemical potential $(\mu_B)$ \cite{PhaseDiag}.

Heavy-ion collisions at RHIC have provided evidence that a
new state of nuclear matter exists at the top RHIC energy 
$\sqrt {s_{NN}}=$200~GeV \cite{BRAHMS,PHOBOS,STAR,PHENIX}.
This new state is characterized by a suppression
of particle production at high $p_T$ 
\cite{suppr1,suppr2,suppr3,suppr4}, a large amount
of elliptic flow $(v_2)$, number-of-constituent-quark
 (NCQ) scaling 
of $v_2$ at intermediate $p_T$
\cite{v2flow} and enhanced correlated yields at large 
$\Delta \eta $ and $\Delta \phi \simeq 0$ (ridge) 
\cite{ridge1,ridge2,ridge3}.

\begin{figure}[h]
\centerline{
\includegraphics[width=6.5cm]{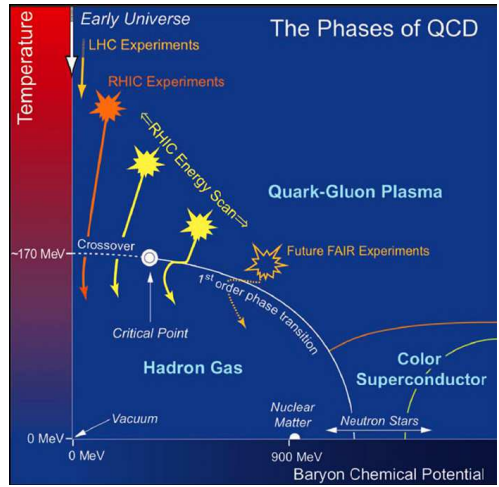}}
\caption{Phase diagram of nuclear matter \protect\cite{BESII}.
\label{f1}}
\end{figure}

Figure \ref{f1} shows a schematic layout of the phases, 
along with hypothesized indications of the regions crossed 
in the early stages of nuclear collisions at various beam energies.

The idea of the Beam Energy Scan (BES) Program at RHIC \cite{BESI}
was to vary the collision energy with aim to search 
for the QCD critical point and QCD phase boundary 
\cite{BESI1,BESI2,BESI3,BESI4,BESI5}. 
The first phase of the Beam Energy Scan program (BES-I) 
along with the existing top energy at RHIC
has allowed the access to a region of the QCD phase diagram (Fig.~\ref{f1}) 
covering a range of baryon chemical potential $(\mu_B)$ from 20 to 420~MeV
corresponding to Au+Au collision energies from $\sqrt {s_{NN}}=$200 to 7.7~GeV, 
respectively. 
Results from BES-I have further supported
the evidence for the quark-gluon plasma (QGP) discovery at the top RHIC energy 
$\sqrt {s_{NN}}=$200~GeV \cite{BESI,BESII}. 
The results of the search for the critical point and the first-order phase
boundary have narrowed the region of interest to collision energies below 
$\sqrt {s_{NN}}=$20~GeV.

The principal challenge remains localization of a critical
point on the QCD phase diagram.
Near the QCD critical point, several thermodynamic properties of the system
such as the heat capacity, compressibility,
correlation length are expected to diverge with a power-law behavior in the
variable $\epsilon  = (T - T_c)/T_c$, where $T_c$ is the critical temperature.
The rate of the  divergence can be described by a set of critical exponents.
 The  critical exponents are universal in the sense that they depend 
 only on  degrees of freedom in the theory and their symmetry, 
 but not on other details of the interactions.
This scaling postulate is the central concept of the theory
of critical phenomena \cite{Stanley1,Stanley2}.
Current lattice QCD calculations suggest that key features of the phase 
diagram like the critical point and the first-order phase transition 
lie within the $\mu_B$ reach of the RHIC BES Phase-II program 
(see \cite{BESII} and refernces therein).

\section{Experiment and data analysis}

The preliminary results presented here are based on data taken
at STAR in Au+Au collisions at $\sqrt {s_{NN}} = $ 7.7, 11.5, 
39, 62.4 and 200~GeV in the year 2010 while 19.6 and 27 GeV were
collected in the year 2011. The transverse momentum spectra 
of negatively charged particle production in the mid-rapidity 
over a wide range of centrality are reported in this paper.

\subsection{STAR detector}

The main detector used to obtain the results on particle spectra
 is the Time Projection Chamber  (TPC) \cite{TPC}.
The TPC is the primary tracking device at STAR and can track up
to 4000 charged particles per event.
It is 4.2 m long and 4 m in diameter.
Its acceptance covers $\pm 1.$ units of pseudo-rapidity
$\eta$ and the full azimuthal angle.
The sensitive
volume of the TPC contains P10 gas (10\% methane,
90\% argon) regulated 
at 2 mbar above atmospheric
pressure. The TPC data are used to determine particle
trajectories, momenta, and particle-type through
ionization energy loss ({\it dE/dx}). STAR's solenoidal
magnetic field used for this low energy Au+Au test
run was 0.5T.
The Time of Flight (ToF) detector \cite{ToF}
(with $2\pi$ azimuthal coverage and $|\eta| < 1.0)$
further enhances the PID capability.
All events were taken with a minimum bias trigger.
The trigger detectors used in these data are the
Beam-Beam-Counter (BBC) and Vertex Position Detector (VPD)\cite{BBC}.
The BBCs are scintillator
annuli mounted around the beam pipe beyond the east
 and west pole-tips of the STAR magnet at
about 375 cm from the center of the nominal interaction
region (IR), and they have an $\eta $ coverage
of $3.8<|\eta|<5.2$ and a full azimuthal ($2\pi$) coverage.
The VPDs are based on the conventional technology
of plastic scintillator read-out by photomultiplier tubes.
They consist of two identical detector
setups very close to the beam pipe, one on each
side at a distance of $|V_z| = 5.6$~m from the center
of the IR.
The details about the design and the other characteristics
of the STAR detector can be found in \cite{TPC}.


All events selected for the analysis   
satisfied the cuts described below.
The primary vertex for each minimum bias event is
determined by finding the best point of common origin
of the tracks measured in the TPC. In order to reject
events which involve interactions with the beam pipe and
 beam-gas interactions, the vertex positions along the 
 $x, y$ and longitudinal ($z$) beam directions 
 are required to be less than 2 cm and 30 cm, respectively.


Track selection criteria for all data sets are presented
in Table~\ref{ta1}. 
Primary tracks are placed on the distance of closest approach 
(DCA) between each track and the event vertex. 
Tracks must have at least 15 points
(nFitPts) used in track fitting. 
 To prevent multiple counting
 of split tracks, at least 52\% of the total possible fit point
are required (nFitPts/nFitPoss). 
A condition is placed on the number of dE/dx points used
 to derive dE/dx  values (ndE/dx). 
 The cut for the $p_T$ ratio of primary and global tracks (pTpr/pTgl)
were used to delete fail tracks. 
The results presented here 
are for transverse momentum $p_T>0.2$~GeV/{\it c} 
and pseudo-rapidity $|\eta| <0.5$ were selected for the analysis.

\begin{table}[t]
{Table 1. Track selection criteria at all energies.}
{\begin{tabular}{@{}cccccc@{}} \toprule
$|\eta|$ & DCA & nFitPts &nFitPts/nFitPoss & ndE/dx & pTpr/pTgl \\ \colrule
$<0.5$   & $<1$~cm & $>15$ & $>0.52$ & $>10$ & $7/10<\& < 10/7$ \\ \botrule
\end{tabular} 
\label{ta1}}
\end{table}

The efficiency is determined by measuring the reconstruction 
efficiency of Monte Carlo tracks 
embedded  in real events and then run through 
the STAR reconstruction chain including  realistic STAR geometry.
For a given collision energy, centrality region,
 and particle species the efficiency is fitted as 
 a function of $p_T$ using the functional form ${\rm eff} = A\cdot exp\{-(B/p_T)^C\}$.
 The parameters  $A$, $B$, and  $C$ are determined for each centrality 
 for unidentified particles using  efficiencies for identified particles.

Figure 2 demonstrates dependence of reconstruction efficiency 
on transverse momentum of particle at energy $\sqrt {s_{NN}}=7.7$
 and 200~GeV and various centralities.  

\begin{figure}[h]
\centerline{
\includegraphics[width=6.3cm]{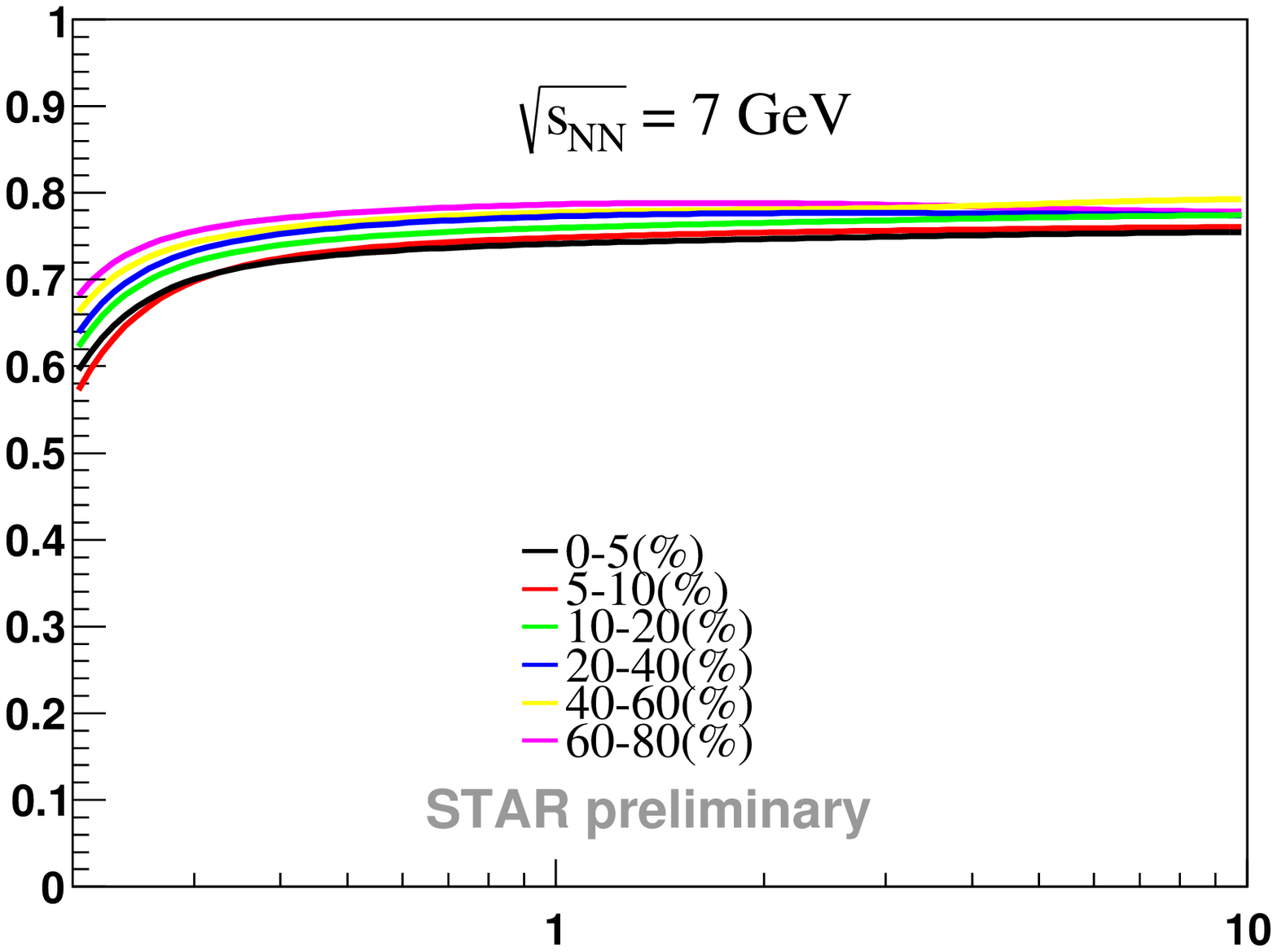}
\hspace*{2mm}
\includegraphics[width=6.0cm]{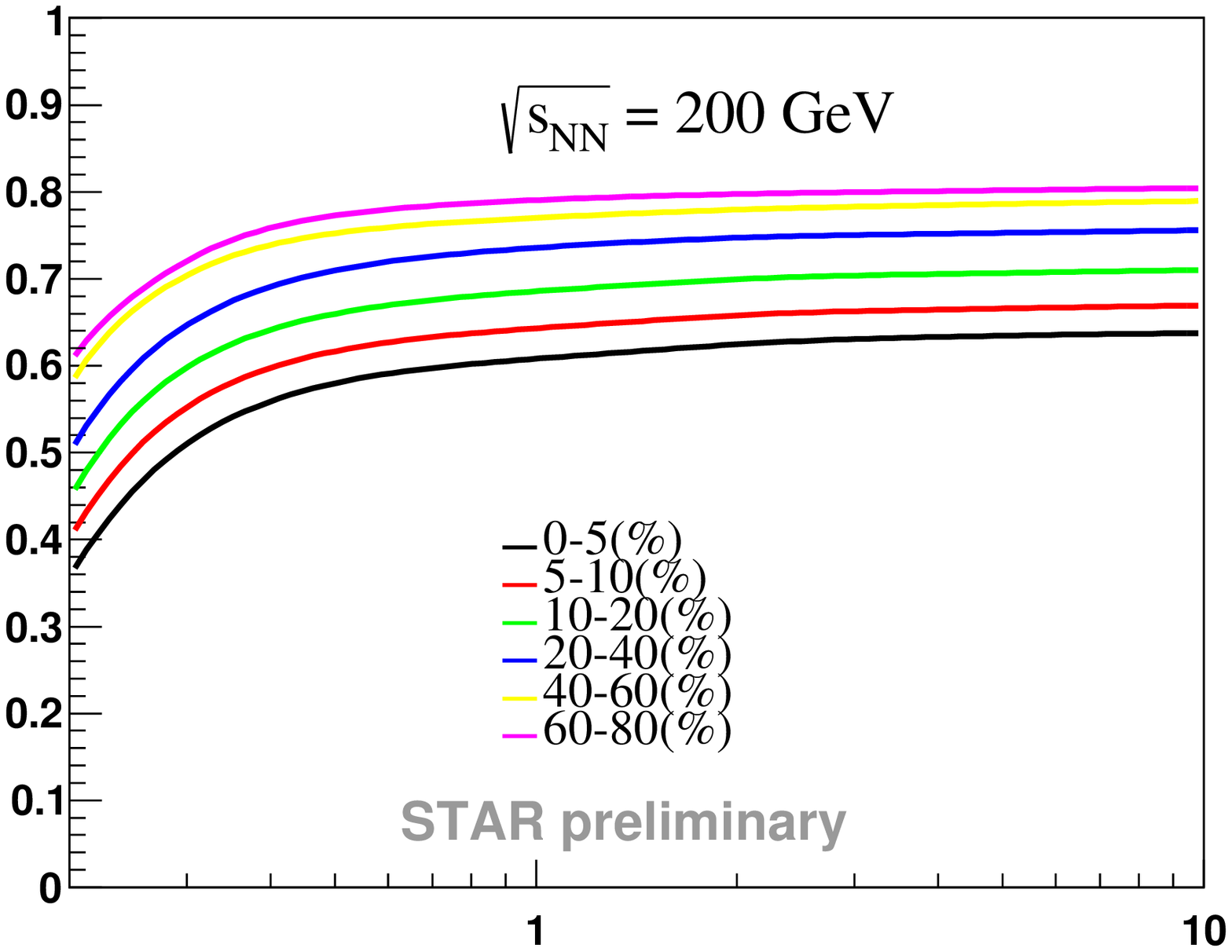}}

\hspace{10mm} a) \hspace{65mm} b)
\caption{Reconstruction efficiency of negatively charged particles produced in Au+Au collisions 
at energies $\sqrt {s_{NN}}= 7.7$ and 200~GeV in the range $|\eta|<0.5$ 
and various centralities  as a function of $p_T$.
\label{f2}}
\end{figure}


Centrality classes in Au+Au collisions at 
$\sqrt {s_{NN}}= 7.7-200$~GeV are defined using 
the number of charged particle tracks 
reconstructed in the main TPC over the full
 azimuth and pseudorapidity $|\eta| < 0.5$. 
 This is generally called   "reference multiplicity (refmult)" 
 in STAR \cite{RefMult1,RefMult2}.
 For each energy, a refmult "correction" is applied on the
 standard definition. This is based on the following steps:
 identifying bad runs, applying acceptance/efficiency 
 correction of refmult for different $z$-vertex positions, and
 performing low refmult correction for trigger inefficiencies
  for different $z$-vertices.
 The refmult distribution is compared/fitted to 
 Monte-Carlo Glauber model simulations.
  The centrality is defined 
 by calculating the fraction of the total cross-section
 obtained from the simulated multiplicity. The events are
 divided into following centrality classes: 
 $0-5\%$, $5-10\%$, $10-20\%$, $20-40\%$, $40-60\%$, $60-80\%$.
Table~\ref{ta2} shows the total number
of events that are used for the analysis for each energy
after the above event selection cuts.
\begin{table}[h]
{
Table 2. Total events analyzed after all analysis cuts for
various energies.}
{\begin{tabular}{@{}ccc@{}} \toprule
$\sqrt {s_{NN}}$ & Year & Events  \\
(GeV) &    & (Millions)  \\ \colrule
7.7   & 2010 & 2    \\
11.5  & 2010 & 7    \\
19.6  & 2011 & 17   \\
27    & 2011 & 33   \\ 
39    & 2010 & 108  \\ 
62.4  & 2010 & 62   \\ 
200   & 2010 & 35  \\ \botrule
\end{tabular}
 \label{ta2}}
\end{table}

\section{Results and discussion} 

\subsection{Transverse momentum spectra}

 The transverse momentum spectrum of
hadrons produced in high-energy collisions of heavy ions
reflects features of constituent interactions
in the nuclear medium.
The medium modification is one of the effects
(recombination, coalescence, energy loss, multiple scattering,...)
 that affects the shape of the spectrum.
 The properties of the created medium
are experimentally studied by variation of the event centrality and collision energy.

\begin{figure}[t]
\vskip 10mm
\centerline{
\includegraphics[width=4.5cm]{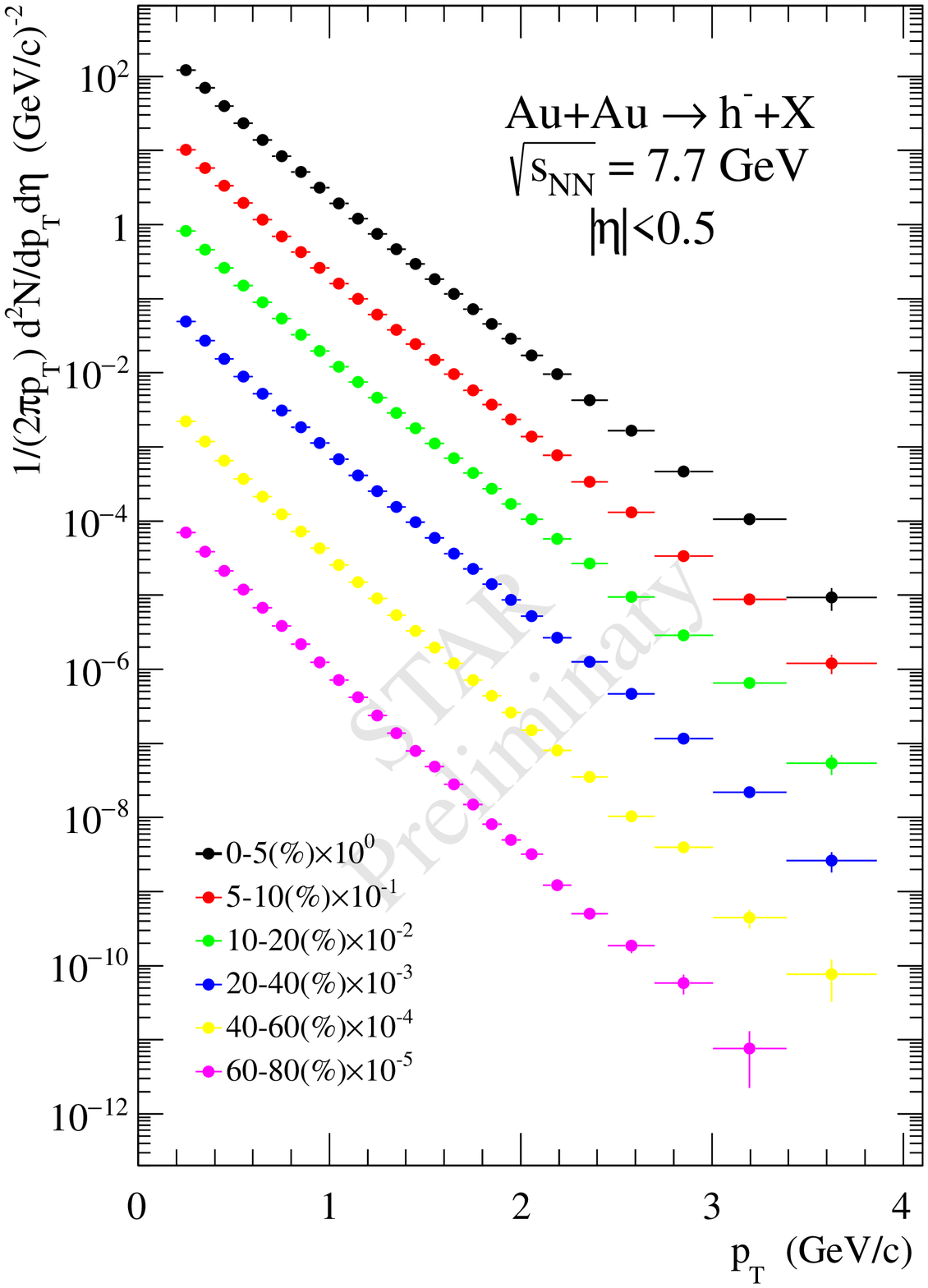}
\hspace*{15mm}
\includegraphics[width=4.5cm]{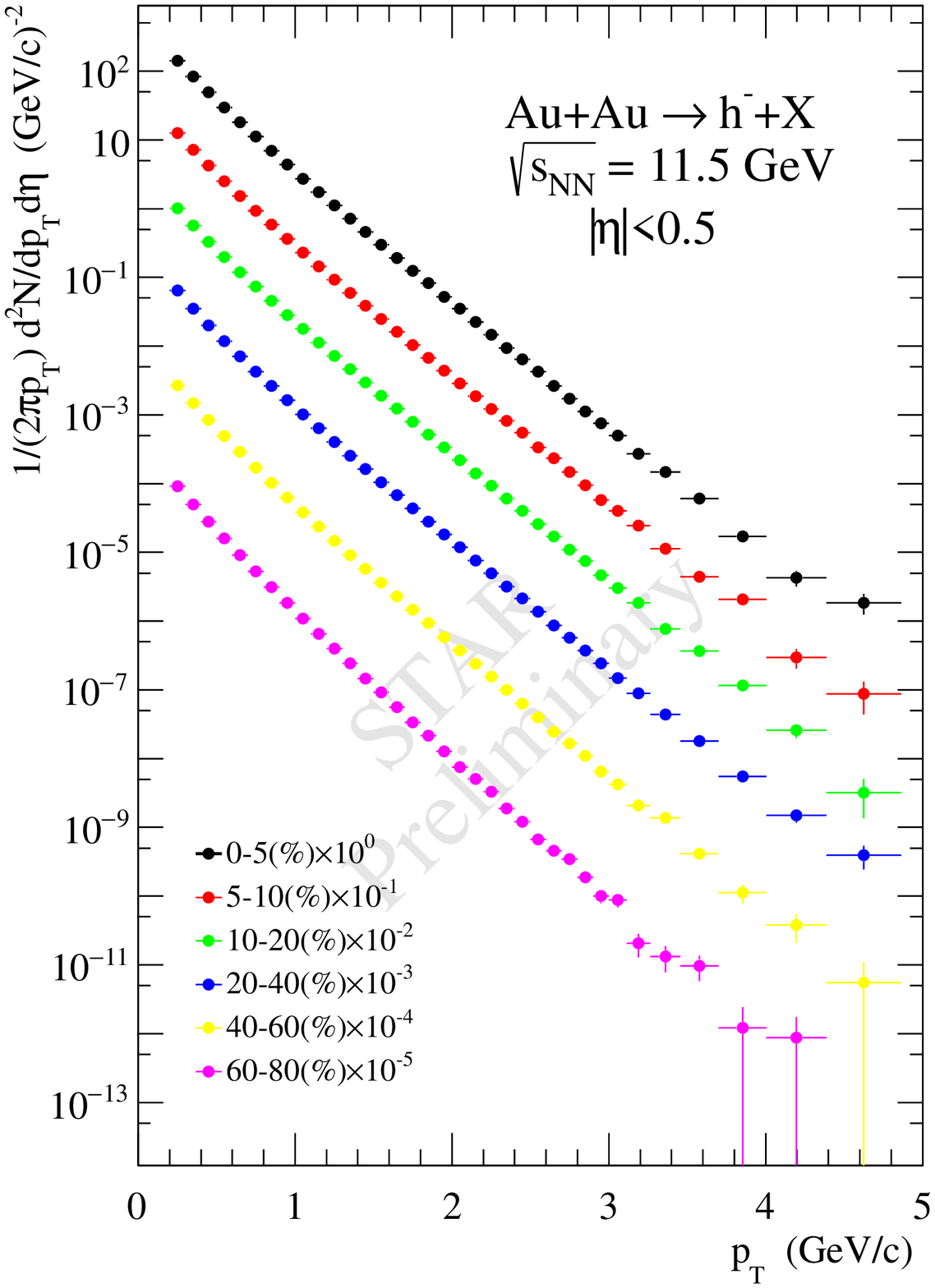}}

\hspace{10mm} a) \hspace{55mm} b)
\vskip 10mm
\centerline{
\includegraphics[width=4.5cm]{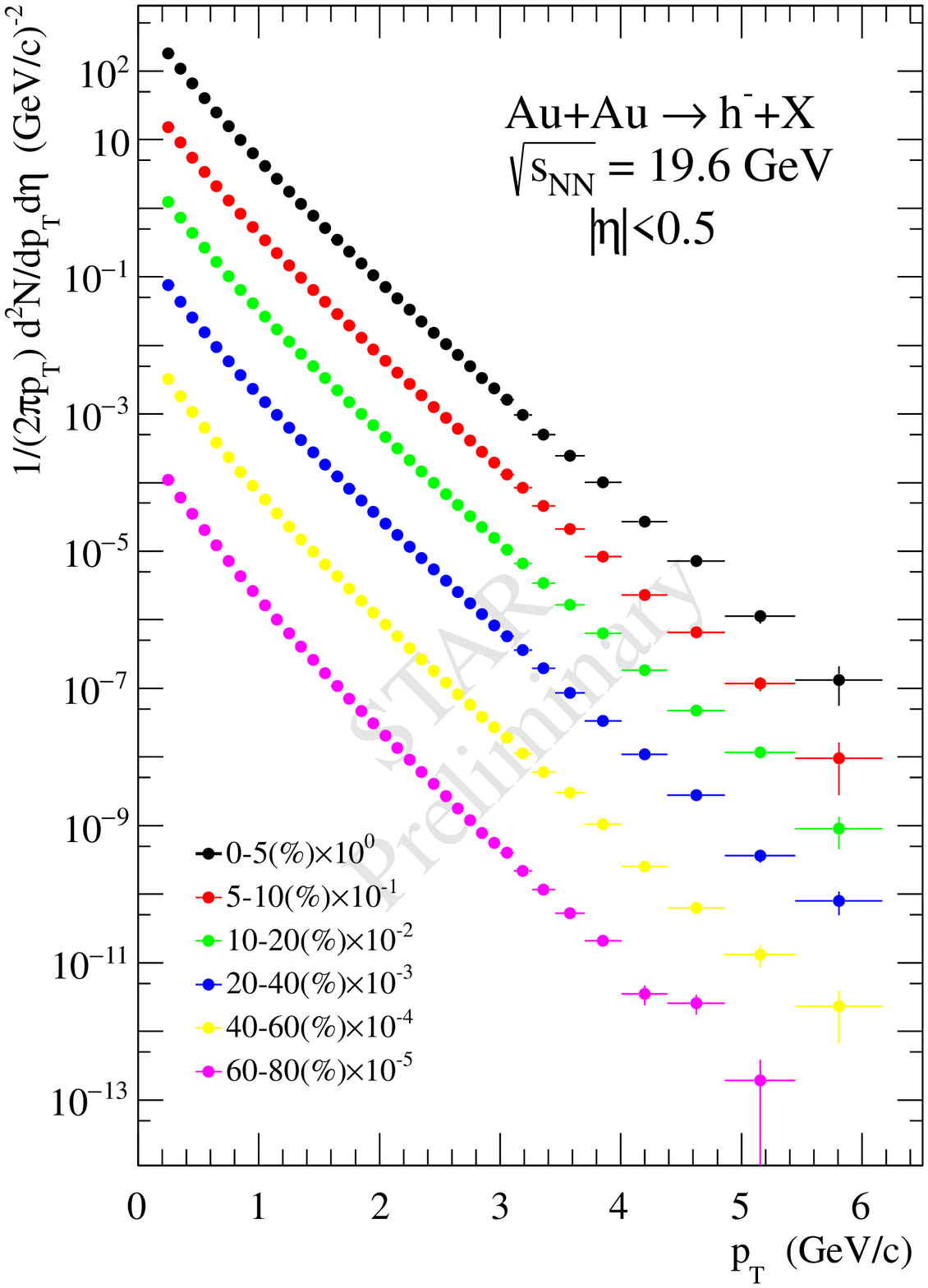}
\hspace*{15mm}
\includegraphics[width=4.5cm]{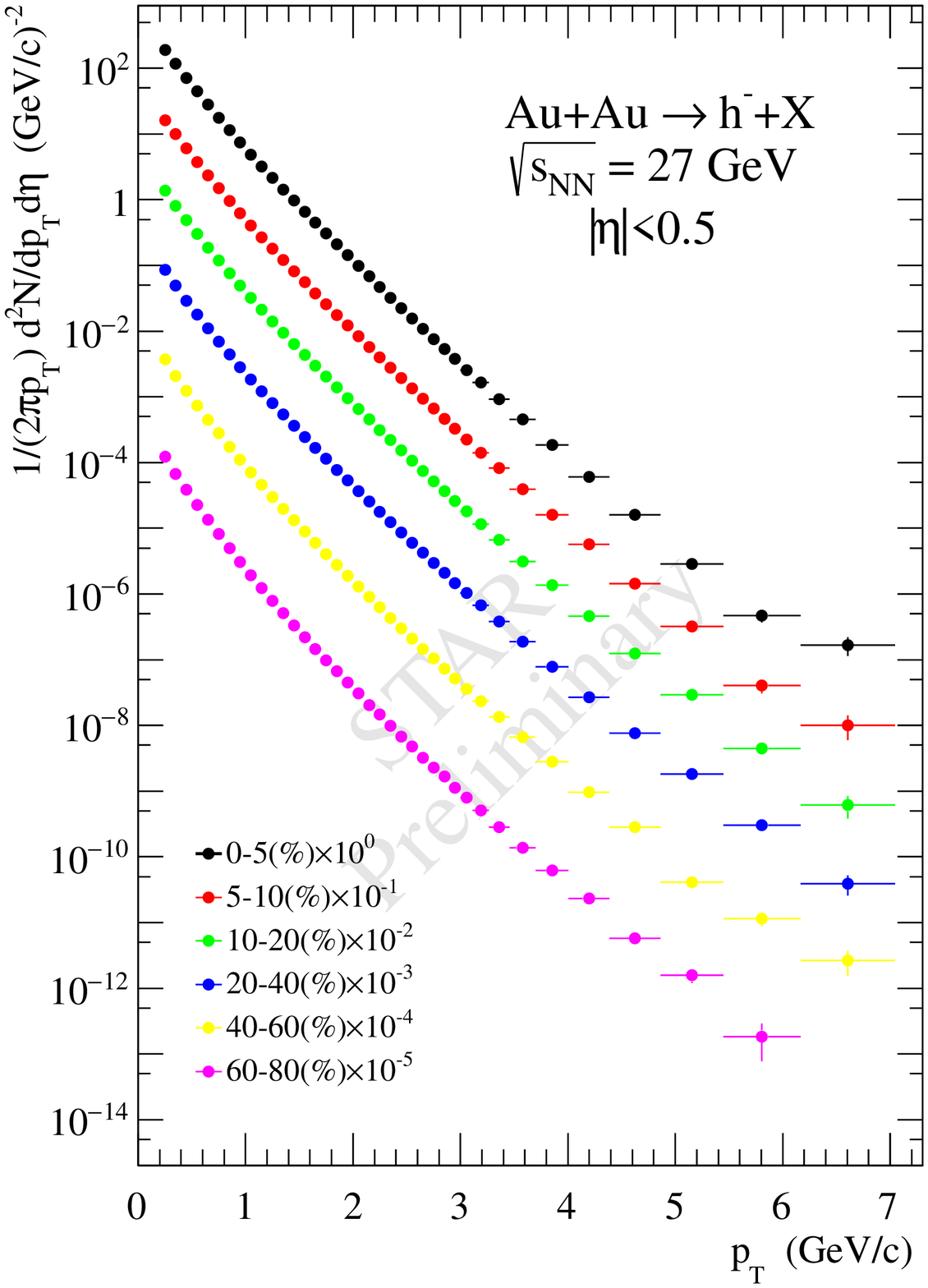}}

\hspace{10mm} c) \hspace{55mm} d)
\caption{Transverse momentum spectra of negatively charged particles produced in Au+Au collisions at energies 
$\sqrt {s_{NN}}= $7.7, 11.5, 19.6 and 
27~GeV in the range $|\eta|<0.5$ and for different centralities
(0-5\%,5-10\%,10-20\%,20-40\%,40-60\%,60-80\%). Errors shown are statistical only.
\label{f3}}
\end{figure}

\begin{figure}[t]
\centerline{
\includegraphics[width=4.5cm]{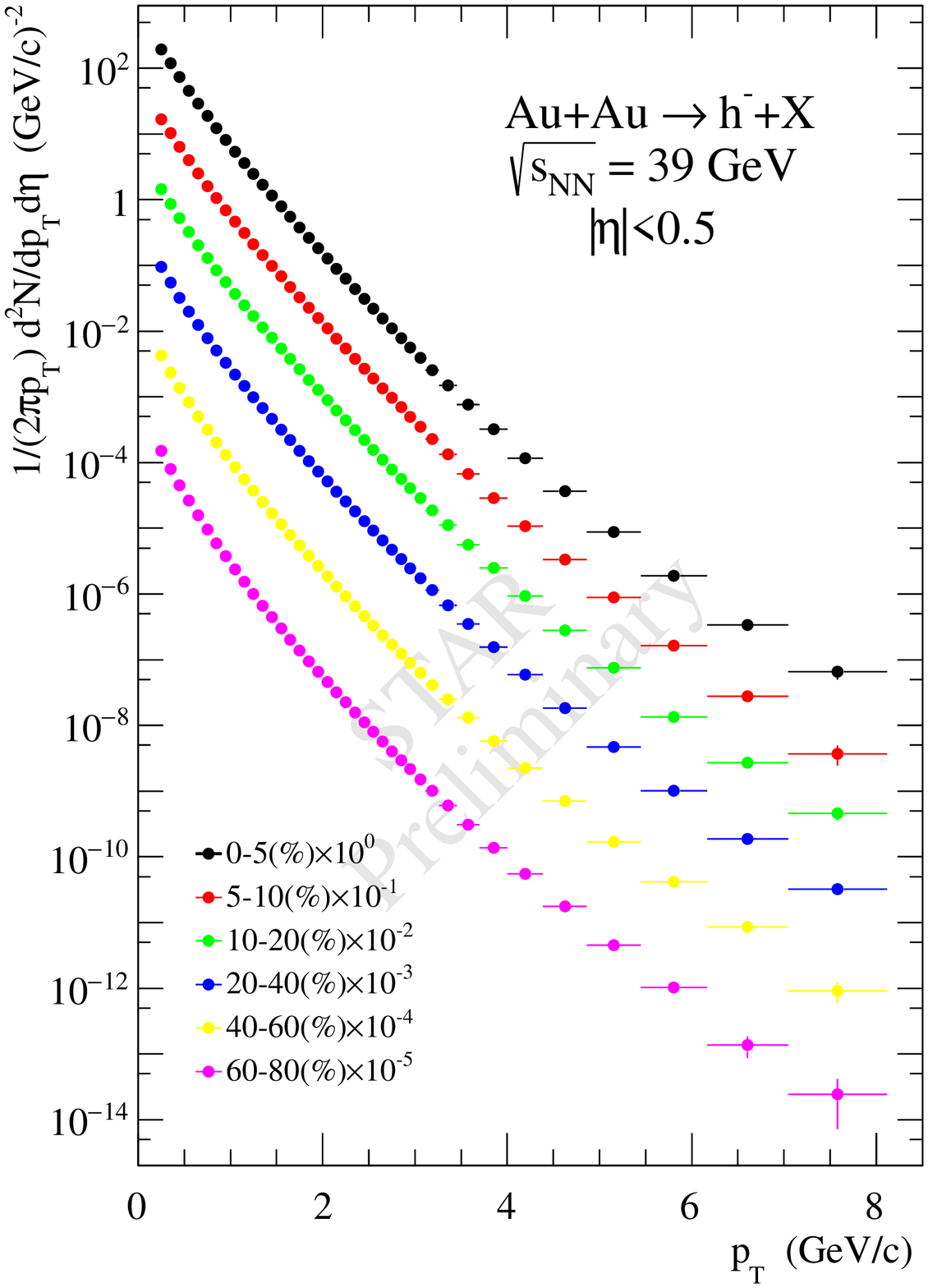}
\hspace*{15mm}
\includegraphics[width=4.5cm]{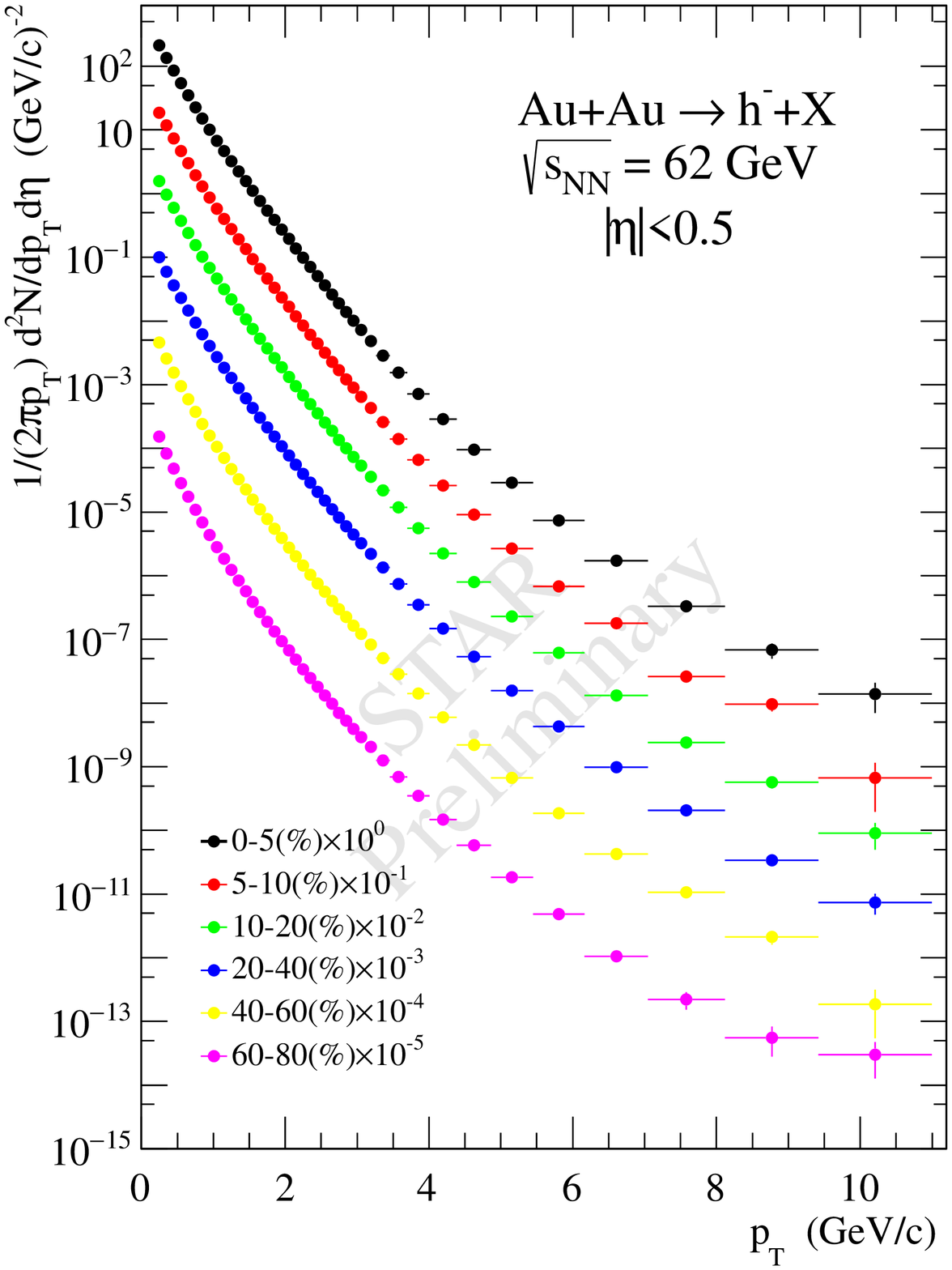}}

\hspace{10mm} a) \hspace{55mm} b)
\vskip 10mm
\centerline{
\includegraphics[width=4.5cm]{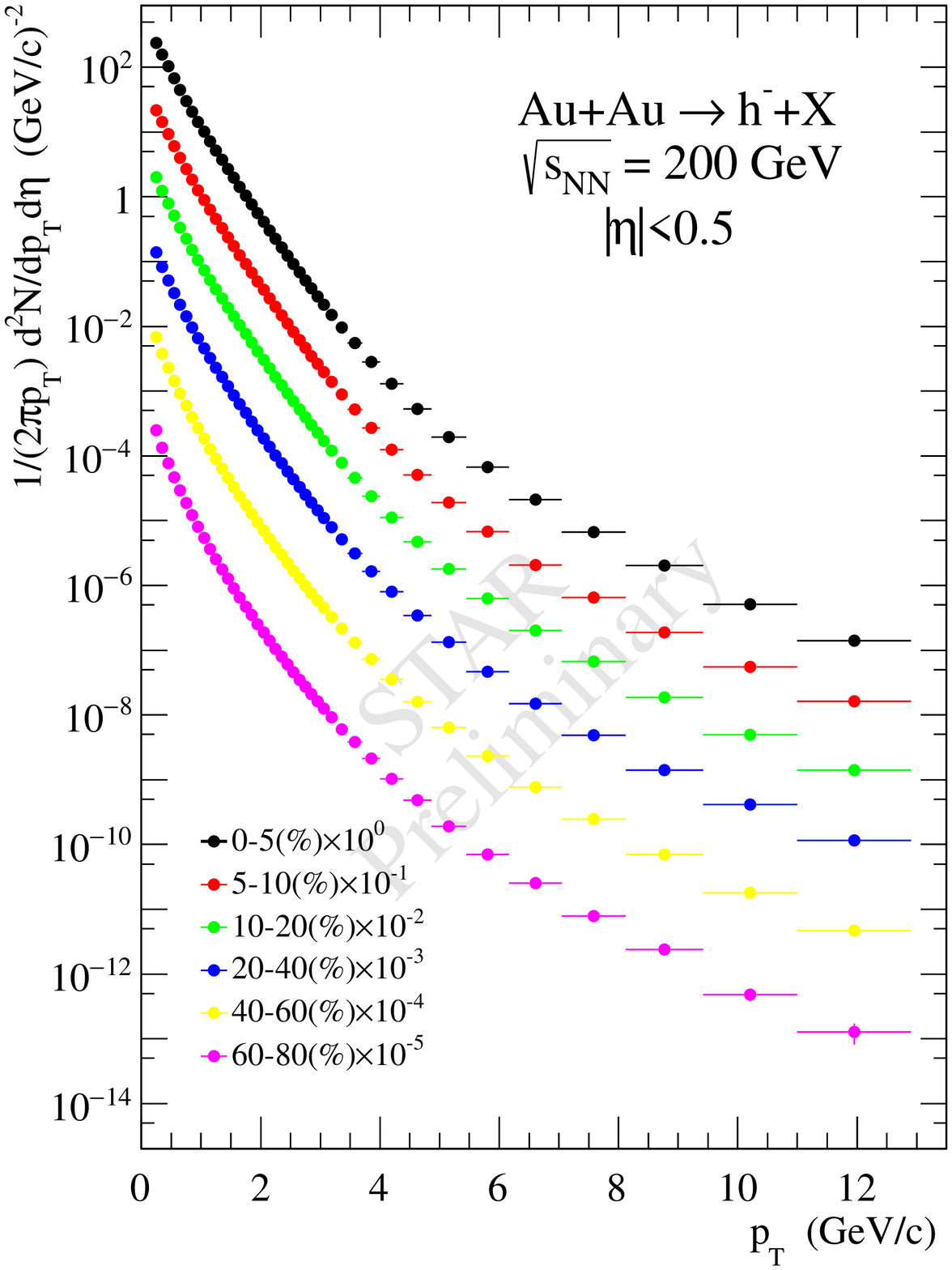}
\hspace*{15mm}
\includegraphics[width=4.5cm]{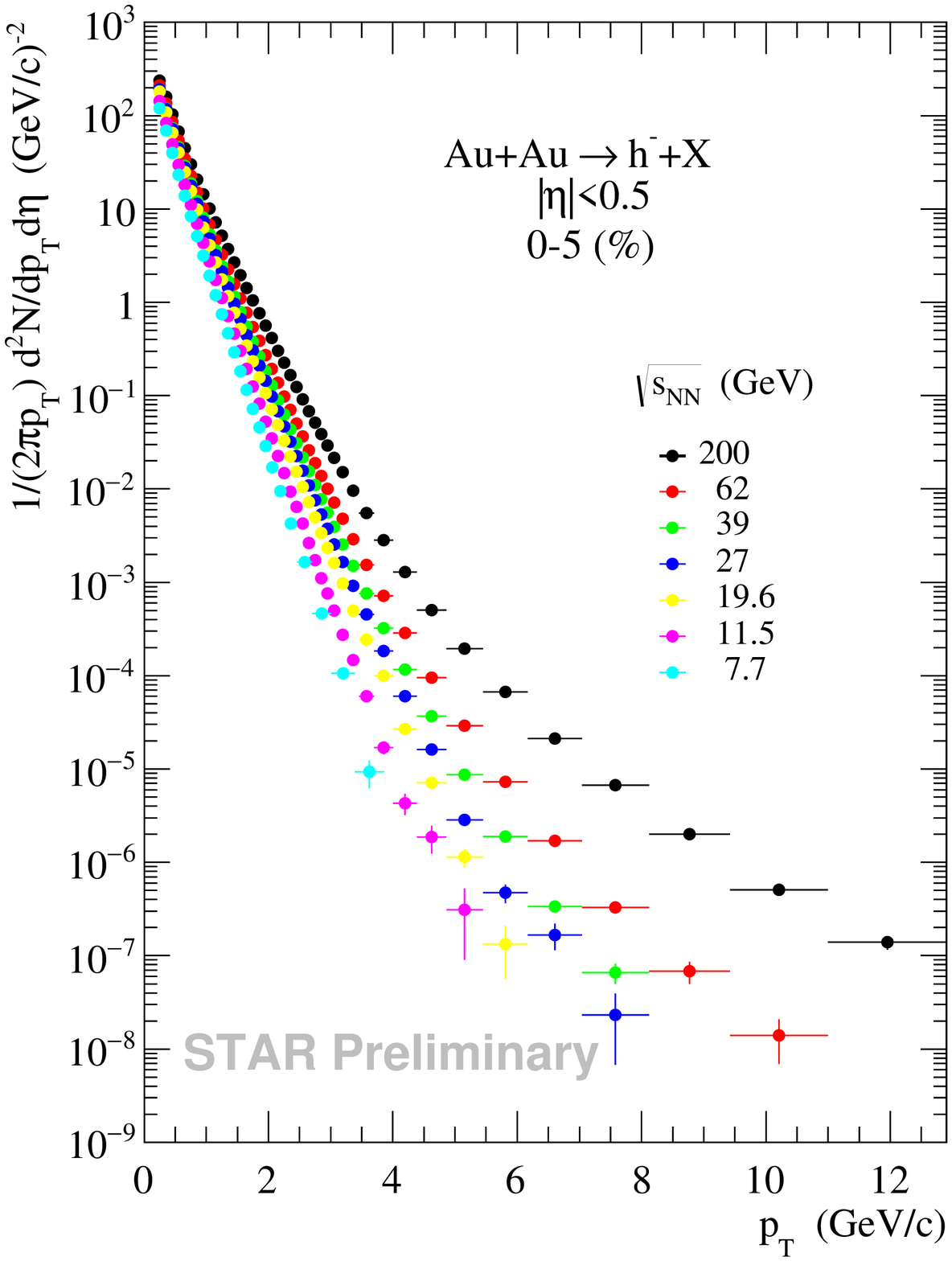}}

\hspace{10mm} c) \hspace{55mm} d) 
\caption{Transverse momentum spectra of negatively charged particles produced in Au+Au collisions at energies 
$\sqrt {s_{NN}}= $39 (a), 62.4 (b) and 200~GeV (c) in the range $|\eta|<0.5$ and for different centralities (0-5\%,5-10\%,10-20\%,20-40\%,40-60\%,60-80\%). 
(d) Spectra for all energies $\sqrt {s_{NN}}= 7.7-200$~GeV  
in central events (0-5\%).
 Errors shown are statistical only.
\label{f4}}
\end{figure}
Figures \ref{f3} and \ref{f4} shows the negatively charged hadron yields
 in Au+Au collisions at
$\sqrt {s_{NN}} = $7.7, 11.5, 19.6, 27, 62.4 and 200~GeV 
at mid-rapidity $|\eta|<0.5$
as a function of transverse momentum $p_T$.
The results are shown for the collision centrality
classes of $0-5\%$, $5-10\%$, $10-20\%$, $20-40\%$, $40-60\%$, $60-80\%$.
The distributions are measured over a wide momentum range
 $0.2<p_T<12.$~GeV/c.
The multiplied factor of 10 is used for visibility.
As seen from Figs. \ref{f3} and \ref{f4} spectra fall more than eight  orders of magnitude.
Spectra demonstrate a strong dependence 
on collision energy and centrality at high $p_T$.  
The exponential behavior of spectra
 at low $p_T$ and the power behavior of spectra 
 at high $p_T$ are observed.  
 Figure \ref{f4}(d) shows  the transverse momentum spectra
of negatively charged particles produced in Au+Au collisions at energies 
$\sqrt {s_{NN}}= 7.7$, 11.5, 19.6, 27, 39, 62.4  
and 200~GeV in the range $|\eta|<0.5$ for most central events ($0-5\%$). 
 The difference in particle yields between different energies 
 increases with transverse momentum
 It is about three orders of magnitude for $0-5\%$ Au+Au 
collisions at $p_T=5$~GeV/c 
between energies  $\sqrt {s_{NN}} = $11.5 and 200~GeV.
As we can see the data presentation in terms of dimensional 
quantities (a particle yield and transverse momentum) 
demonstrates a strong dependence on the scale.

\subsection{Self-similarity of negative particle production}

The study of critical phenomena in nuclear matter is most preferable 
in terms of dimensionless variables.
The idea of the Beam Energy Scan programs at RHIC  is to vary the collision energy 
and look for the signatures of QCD phase boundary and QCD critical point 
i.e. to span the phase diagram from the top RHIC energy to the lowest possible energy.
 Turn-off of signatures of QGP established at $\sqrt {s_{NN}}=$200~GeV 
 would suggest the crossing of phase boundary. 

Self-similarity of hadron interactions is assumed to be a very fruitful concept 
to study collective phenomena in the hadron and nuclear matter. 
Important manifestation of
 this concept is a notion of scaling itself. 
Scaling in general means self-similarity at different scales. 
  The approach based on $z$-scaling suggected in \cite{Zscal1,Zscal2,ZscalAA} 
  is treated as manifestation of the self-similarity property of the structure 
 of colliding objects (hadrons, nuclei), the interaction mechanism 
 of their constituents, and the process of constituent fragmentation 
 into real hadrons. We use the approach for analysis of the obtained spectra
 (see Figs.~\ref{f3},\ref{f4}). 
   The approach relies on a hypothesis about self-similarity
of hadron interactions at a constituent level.
The assumption of self-similarity  transforms to the requirement
of simultaneous description of transverse momentum spectra corresponding
to different collision energies, rapidities, and centralities
by the same scaling function $\psi(z)$ depending on a similarity parameter $z$.
The scaling function for the collision of nuclei is expressed in terms of the experimentally 
measured inclusive invariant cross section  ($E {{d^3\sigma}/{dp^3}}$), 
the multiplicity density ($dN/d\eta$), 
and the total inelastic cross section ($\sigma_{\rm in}$), as follows:
\begin{equation}
\psi(z) = -{ { \pi} \over { (dN/d\eta)\ \sigma_{\rm in}} }
J^{-1} E {{d^3\sigma} \over {dp^3}}.
\label{eq:r5}
\end{equation}
Here  $J$ is the Jacobian for the transformation
from $\{ p_T^2, y \}$ to $\{ z, \eta \}$. The function $\psi$
is interpreted as a probability density
to produce an inclusive particle with the corresponding value of $z$.
The similarity parameter $z$ is expressed via
momentum fractions $(x_1, x_2, y_a, y_b)$,
multiplicity density, and parameters $\delta_{1,2}, \epsilon_{a,b}, c$, as follows:
\begin{equation}
z =z_0 \Omega^{-1},
\label{eq:r8}
\end{equation}
where
$ z_0 = {  \sqrt {s_{\bot}}    } / [{(dN_{ch}/d\eta|_0)^c  m_N]},\ \ 
\Omega =
(1-x_1)^{\delta_1}(1-x_2)^{\delta_2}(1-y_a)^{\epsilon_a}(1-y_b)^{\epsilon_b}.$
Here $m_N$ is a nucleon mass. The quantity $\sqrt {s_{\bot}}$ is the transverse kinetic energy of the binary subprocess
consumed on the production of the inclusive particle with mass $m_1$ and
its counterpart with mass $m_2$.
The constituents of the incoming nuclei carry fractions $x_1, x_2$
of their momenta.
The inclusive particle carries the momentum fraction $y_a$
of the scattered constituent.
The parameters $\delta_{1,2}$  and  $\epsilon_{a,b}$ describe
 structure of the colliding nuclei
 and fragmentation process, respectively.
The parameter $c$ is interpreted
as a "specific heat" of the created medium.
Simultaneous description of different spectra with the same
$\psi(z)$ puts strong constraints on the values of these parameters,
  and thus allows for their determination.
It was found that $\delta_{1,2}$ and $c$ are constant 
at high energies, $\epsilon_{a,b}$ depends on multiplicity and $m_1=m_2\equiv m$. 
For the obtained values of $\delta_{1,2}, \epsilon_{a,b}$ and $c$,
the momentum fractions are determined to minimize the resolution
$\Omega^{-1}(x_1,x_2, y_a, y_b) $,
which enters in the definition of the variable $z$.
The system of the equations
$\partial \Omega/\partial x_1 = 
\partial \Omega/\partial x_2 =
\partial \Omega/\partial y_a =
\partial \Omega/\partial y_b =0$
was numerically resolved under the constraint
$(x_1P_1+x_2P_2-p/y_a)^2=(x_1 M_1+x_2 M_2+m/y_b)^2$, which
has sense of the momentum conservation law of a constituent subprocess \cite{Zscal1,Zscal2,ZscalAA}. 
The extended version for the approach was applied for description of hadron yields 
in nucleus-nucleus collisions \cite{ZscalAA}. 
The fractal dimension of the nucleus is expressed in terms 
of the nucleon fractal dimension $(\delta)$  and the atomic number of nucleus $(A)$
as follows, $\delta_A= A\delta $.
 The fragmentation fractal dimension 
in nucleus-nucleus collisions is parameterized in the form:
$\epsilon_{AA}=\epsilon_0(dN/d\eta)+\epsilon_{pp}$. 
It depends on multiplicity density and nucleon 
fragmentation dimension ($\epsilon_{pp}=0.2$). Here  $\epsilon_0$
is a fitting parameter.
In this approach the energy loss of the scattered
constituent during its fragmentation in the inclusive particle
is proportional to the value $(1-y_a)$. 
 We would like to emphasize once more that dimensionless quantities, like 
 momentum fractions, a scaling function and similarity parameter,
 are expressed via dimensional experimental measurable quantities.
 
\begin{figure}[t]
\centerline{
\includegraphics[width=6.3cm,height=6.5cm]{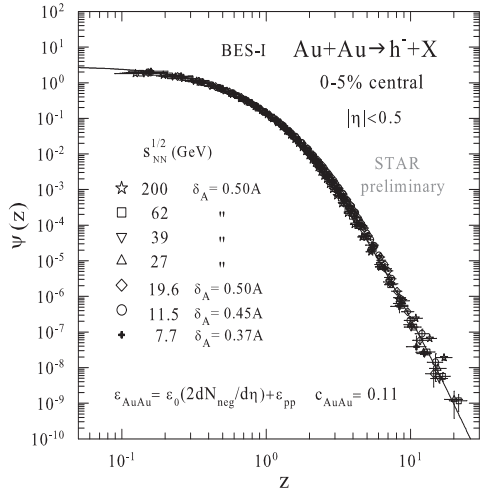}}
\caption{
Scaling function $\psi(z)$ as a function of self-similarity parameter $z$ for
corresponding transverse momentum spectra of negatively charged particles produced
 in Au+Au collisions at energies 
 $\sqrt {s_{NN}}= 7.7-200$~GeV in the range $|\eta|<0.5$
 for most central events (0-5\%).
\label{f5}}
\end{figure}
 
Figure \ref{f5} shows scaling function $\psi(z)$ as a function 
of self-similarity parameter $z$.
The data $z$ presentation 
corresponds to spectra of negatively charged particles produced 
in Au+Au collisions at energies $\sqrt {s_{NN}} = 7.7-200$~GeV 
in the range $|\eta|<0.5$ for most central events (0-5\%) 
as a function of the transverse momentum. 
We observe "collapse" of the data onto a single curve.
The solid line is a fitting curve for these data. 
The derived representation shows the universality of the shape 
of the scaling curve for the BES energies. 
The found regularity (the shape of the function  
and its scaling behavior in the wide kinematic range)
can be treated as manifestation of the self-similarity 
of the structure of colliding objects, 
interaction mechanism of their constituents, and processes 
of fragmentation of these constituents into real registered particles. 
As we can see from Fig.~\ref{f5}, the scaling function exhibits two kinds 
of behavior: 
one in the low-$z$ and the other in the high-$z$ region. 
The first regime corresponds to saturation 
of the scaling function with the typical flattening out. 
The second one corresponds to the power behavior
of the scaling function pointing on self-similarity 
in constituent interactions at small scales.

\begin{figure}[t]
\centerline{
\includegraphics[width=6.cm]{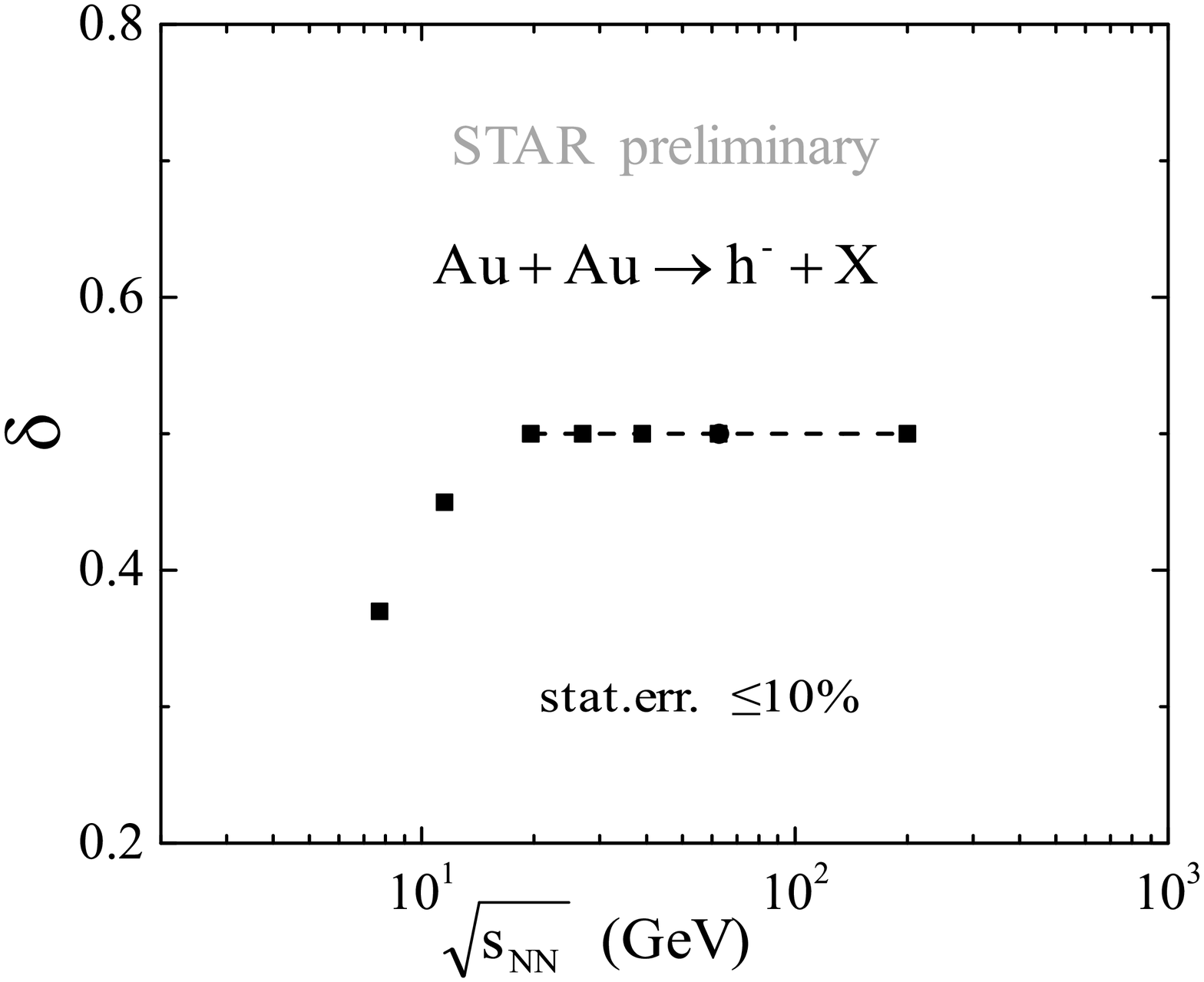}
\hspace*{5mm}
\includegraphics[width=5.9cm]{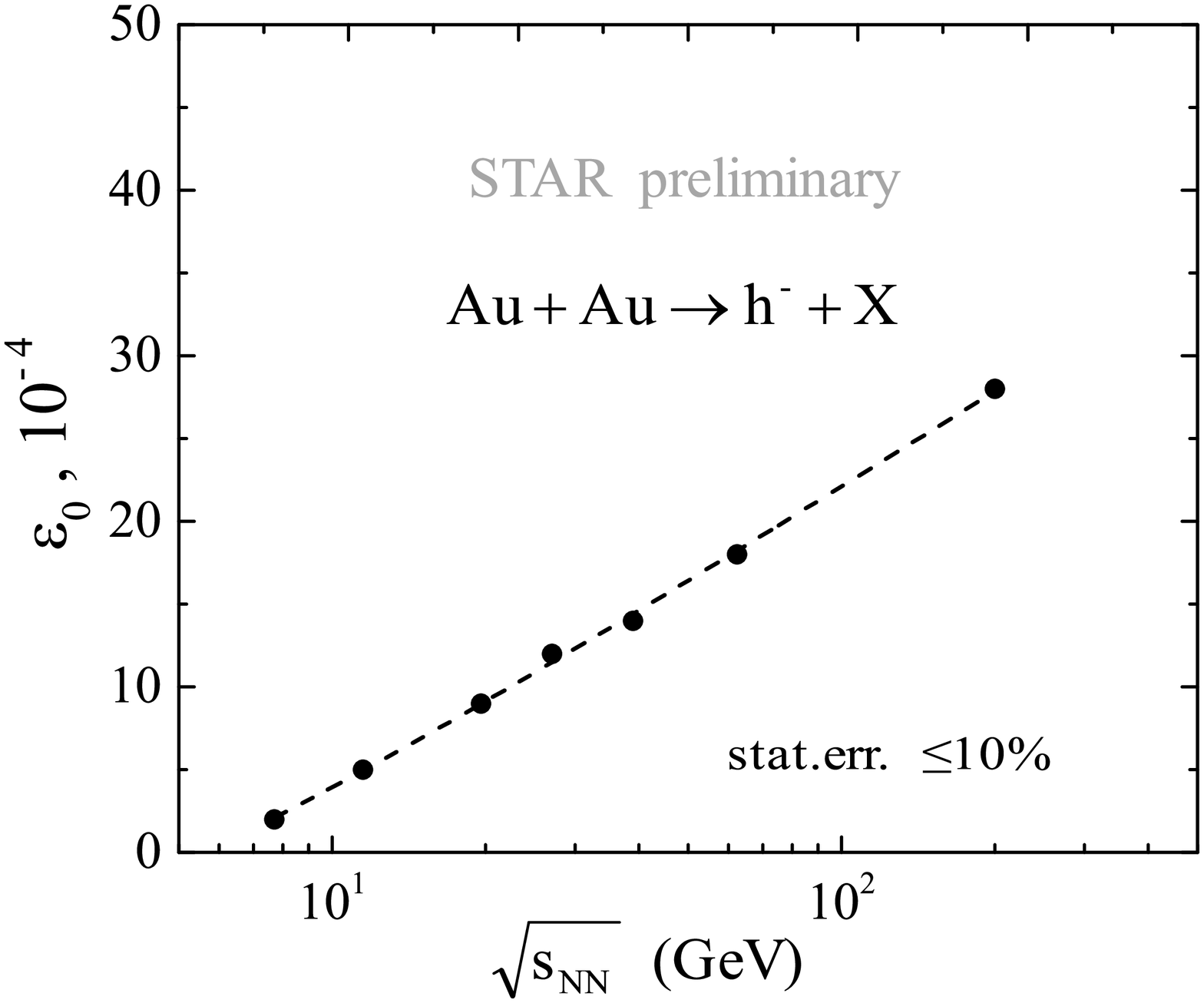}}

\hspace{10mm} a) \hspace{60mm} b)
\caption{Model parameters $\delta$ and $\epsilon_{0}$ 
as a function of collision energy $\sqrt {s_{NN}}$. 
\label{f6}}
\end{figure}

\subsection{Model parameters}

The model parameters - nucleus fractal dimension $\delta_A$, 
fragmentation fractal dimension $\epsilon_{AA}$, and "heat capacity" $c$, 
are determined from the requirement of scaling behavior of $\psi(z) $ 
(see Fig.~\ref{f5}) as a function of self-similarity parameter $z$.  
The fractal dimension of the nucleus and 
the  fragmentation dimension are equal to  $\delta_A= A \delta$
and  $\epsilon_{AA}=\epsilon_0(dN/d\eta)+\epsilon_{pp}$, respectively.
Figure \ref{f6} shows the dependence of these parameters on the collision energy. 
It was found that $\delta $ is independent of the energy 
 for $\sqrt {s_{NN}} \ge 20$~GeV, $\epsilon_{0}$ increases with the energy 
and $c$ is independent of the energy over the range $\sqrt {s_{NN}} =7.7-200$~GeV.  
Note that, while $\delta $ decreases with decreased energy
for $\sqrt {s_{NN}} \le 20$~GeV, the function $\psi(z)$ preserves 
the shape in this range.

\subsection{Constituent energy loss}

The nuclear modification factor $R_{AA}$ measured
at RHIC at $\sqrt {s_{NN}}=62.4,130$ and 200~GeV
strongly shows a suppression of the charged hadron spectra
at $p_T>4$~GeV/c.
   This suppression \cite{suppr1,suppr2,suppr3,suppr4} 
  was one of the first indications of a strong final-state
modification of particle production in Au+Au collisions
that is now generally ascribed to energy loss
of the fragmenting parton in the hot and dense medium.
At the BES-I energies the nuclear modification factor  
drastically changes with a collision energy \cite{BESI,BESII}. 
It increases with transverse momentum  as the energy decreases.
The study of the evolution of the energy loss with collision
energy has relevance to the evolution of created nuclear
matter, and can be useful for searching for signature
of phase transition and a critical point.
The measured spectra (see Fig.~\ref{f3},\ref{f4}) allow us to estimate
constituent energy loss in charged hadron production
in Au+Au collisions at BES energies. The  estimations 
are based on a microscopic scenario of particle production  
proposed in \cite{Zscal1,Zscal2,ZscalAA}.
The energy loss of the scattered constituent during 
its fragmentation in the inclusive particle
is proportional to the value $(1-y_a)$. 

\begin{figure}[t]
\centerline{
\includegraphics[width=5.5cm]{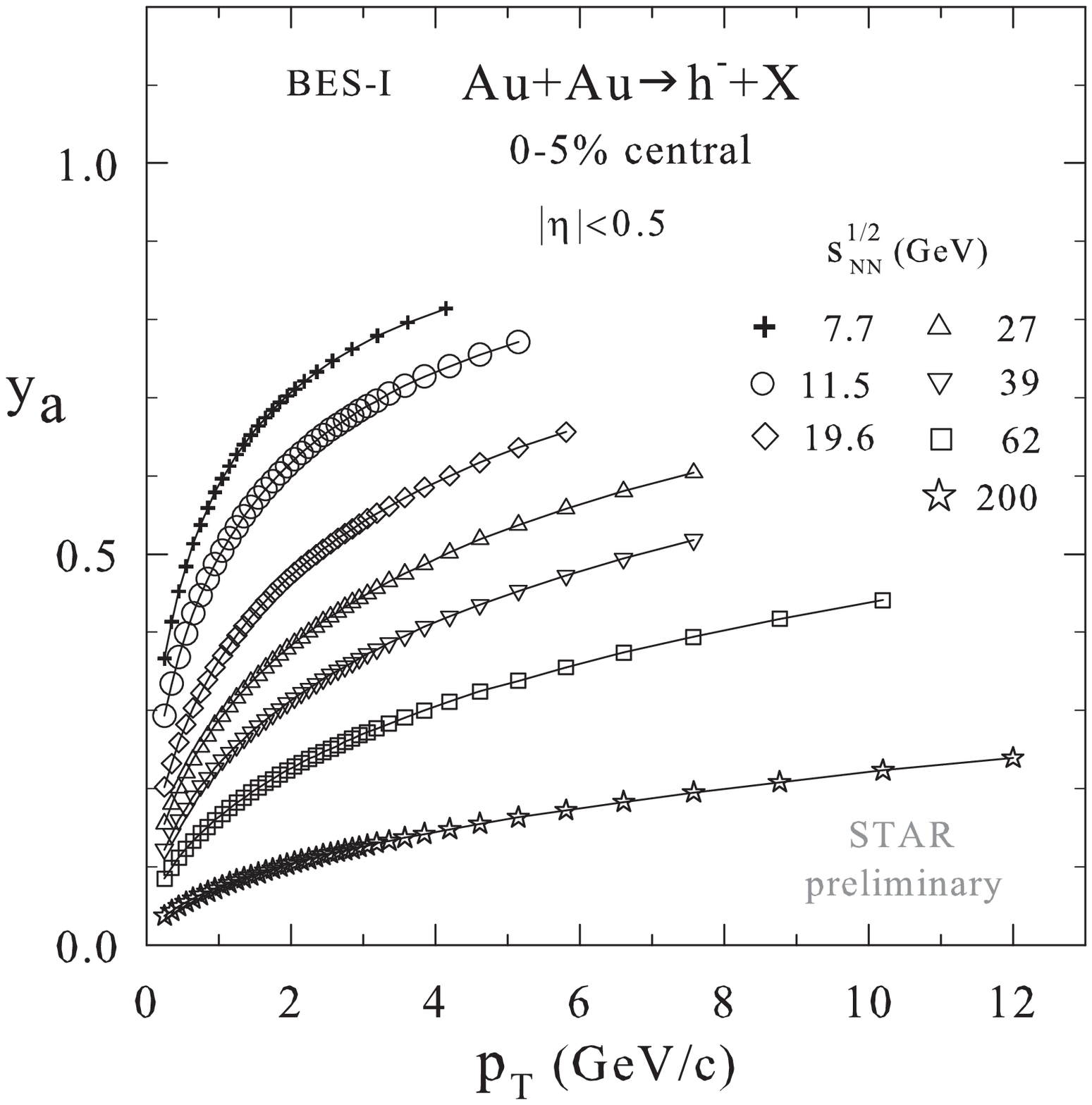}
\hspace*{5mm}
\includegraphics[width=5.5cm]{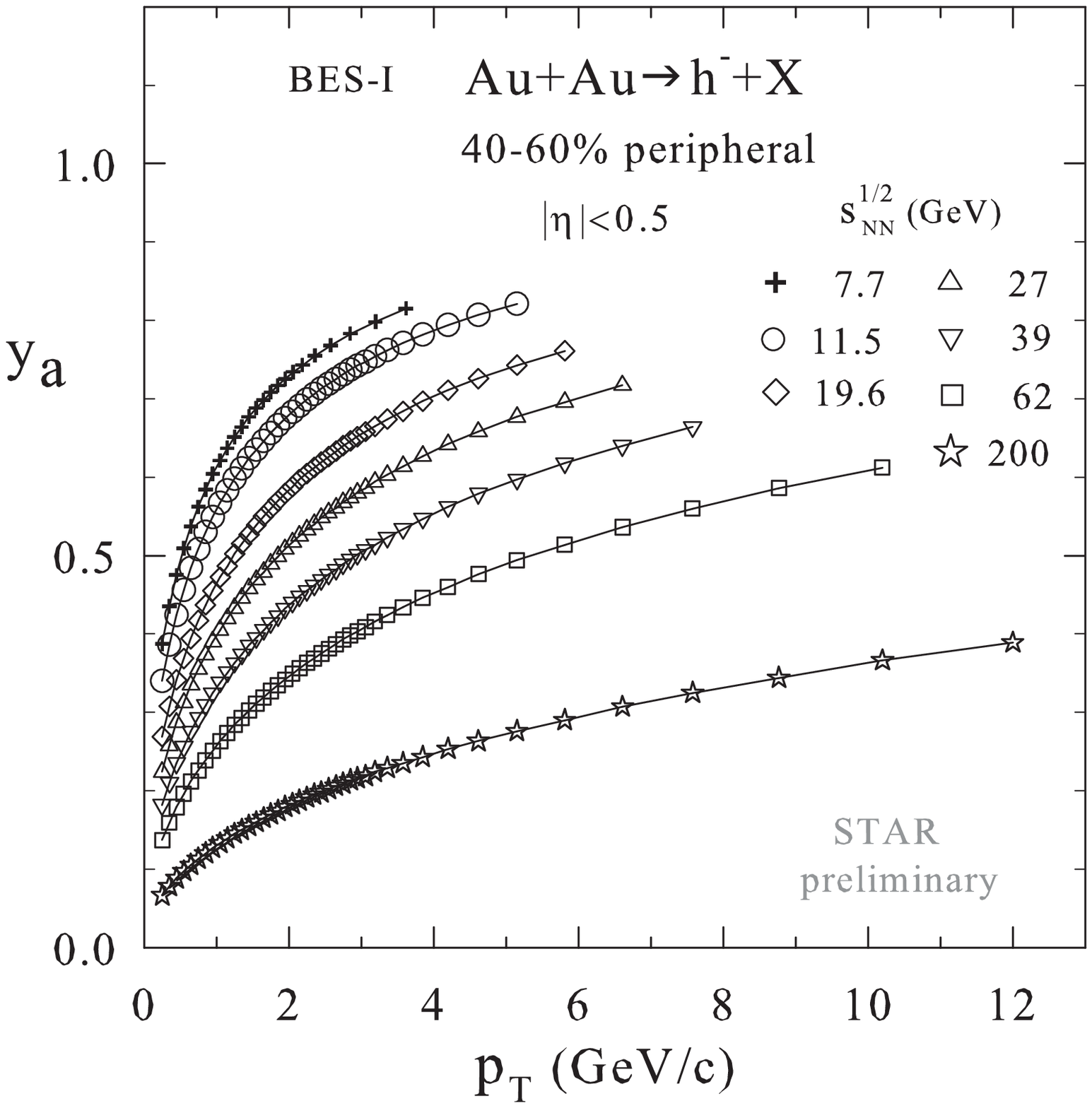}}

\hspace{10mm} a) \hspace{55mm} b)
\caption{Momentum fraction $y_a$  for negatively charged particle production
in most central ($0-5\%$) and peripheral ($40-60\%$) 
\  Au+Au collisions at mid-rapidity ($|\eta|<0.5$) 
as a function of the energy $\sqrt {s_{NN}}$ and  hadron transverse momentum $p_T$.
\label{f7}}
\end{figure}
Figure \ref{f7} shows the dependence of the fraction
$y_a$ for central ($0-5\%$) and peripheral  ($40-60\%$) 
\  Au+Au collisions 
on the transverse momentum over a range $\sqrt {s_{NN}}=7.7-200$~GeV.
The behavior of $y_a$ demonstrates a monotonic growth with $p_T$.
It means that the energy loss associated with the production of
a high-$p_T$ hadron is smaller than that
with lower transverse momenta.
The decrease of $y_a$ with 
collision centrality represents
larger energy loss in the central
collisions as compared with peripheral interactions.
The energy dissipation grows as the collision energy increases.
It is estimated to be about $20\%$ at $\sqrt {s_{NN}}=7.7$~GeV and
$85\%$  at $\sqrt {s_{NN}}=200$~GeV  for $p_T\simeq 4$~GeV/c 
for central events ($0-5\%$), respectively.
These estimations are in agreement with the energy dependence 
of the model parameters (see Fig.~\ref{f6}).
Discontinuity of these parameters was assumed to be a signature
of phase transition and a critical point.

\section{Summary}

In summary, we have presented the first STAR results for
negatively charged hadron production in Au+Au collisions 
 at $\sqrt {s_{NN}} =  7.7-200$~GeV from 
 the Beam Energy Scan program at RHIC. 
The spectra of particle yields at mid-rapidity were measured over 
a wide range of transverse momentum.
The spectra demonstrate a strong dependence on 
the collision energy  and centrality which enhances with $p_T$. 
An indication of self-similarity of negatively charged particle 
production in Au+Au collisions was found. 
It is seen in the presentation of spectra  by scaling function $\psi$
in dependence on self-similarity parameter $z$ for all energies.

The obtained spectra of negatively charged  hadrons in Au+Au collisions
were used to estimate of a constituent energy loss
as a function of energy and centrality of collisions and transverse momentum 
of inclusive particle in the $z$-scaling approach.
It was shown that the energy loss increases with the
collision energy and centrality, and decreases with $p_T$.

The energy dependence of the model parameters - the fractal 
and fragmentation dimensions and "specific heat",  was studied.
All parameters demonstrate smooth behavior with energy for most central events. 
It is expected that the second phase of BES will allow us 
to clarify the existence of a discontinuity and  strong correlation
of these parameters with much higher statistics.
 
The presented results provide an additional motivation
for the Beam Energy Scan-II at the STAR experiment \cite{BESI,BESII}.
The large and uniform acceptance and extended particle identification is suitable
for a critical point search. 
The BES Phase-II program will allow for high-statistics measurements,
with an extended kinematic range in rapidity and transverse momentum,
using sensitive observables, to reveal the structure 
of the QCD phase diagram \cite{BESII}.


\end{document}